\documentclass{article}
\usepackage{spconf}

\usepackage{makecell}

\usepackage{cite}
\usepackage{amsmath,amssymb,amsfonts}
\usepackage{algorithmic}
\usepackage{graphicx}
\usepackage{textcomp}
\usepackage{xcolor}

\usepackage{enumitem}
\usepackage{array}
\usepackage{mathtools}
\usepackage{tcolorbox}
\usepackage{color}
\usepackage{theorem}
\usepackage{amssymb}
\usepackage{caption}
\usepackage{subcaption}
\usepackage{cite,hyperref}
\usepackage{cases}
\usepackage{url}
\usepackage{algorithmic}
\usepackage{algorithm}
\usepackage{tikz}
\usetikzlibrary{shapes,arrows}
\input{my_styles.sty}

\usepackage{multirow}
\usepackage{hhline}

\newcommand{\dpgap}{\Delta\mathrm{DP}}
\newcommand{\dpngap}{\Delta\mathrm{DP}_{\mathrm{node}}}

\usepackage{moresize}
\usepackage{pgfplots}
\pgfplotsset{compat=1.17}
\pgfplotstableset{col sep=comma}
\newtcolorbox{myblockt}[1]{colback=urblue!5!white,
	colframe=urblue,fonttitle=\bfseries,
	title=#1}

\newtcolorbox{myblock}{colback=urblue!5!white,
	colframe=urblue,fonttitle=\bfseries}

\def\BibTeX{{\rm B\kern-.05em{\sc i\kern-.025em b}\kern-.08em
    T\kern-.1667em\lower.7ex\hbox{E}\kern-.125emX}}

\begin{document}

\tikzset{every mark/.append style={scale=1.5, solid}, font=\footnotesize}
\pgfplotsset{
    width=1.05\textwidth,
    legend style={
        font=\ssmall ,  
        inner xsep=1pt,
        inner ysep=1pt,
        nodes={inner sep=1pt}},
    legend cell align=left,
	every axis/.append style={line width=0.5pt},
	every axis plot/.append style={line width=1.25pt},
    every axis y label/.append style={yshift=-5pt}
}

\title{Mitigating subpopulation bias for fair network topology inference}
\name{Madeline Navarro$^\dagger$, Samuel Rey$^*$, Andrei Buciulea$^*$, Antonio G. Marques$^*$, Santiago Segarra$^\dagger$
\thanks{
This work was partially supported by NSF under award CCF-2008555, Spanish AEI Grants PID2019-105032GB-I00, TED2021-130347B-I00, PID2022-136887NB-I00, and by the Autonomous Community of Madrid within the ELLIS Unit Madrid framework.
Research was sponsored by the Army Research Office and was accomplished under Grant Number W911NF-17-S-0002. The views and conclusions contained in this document are those of the authors and should not be interpreted as representing the official policies, either expressed or implied, of the Army Research Office or the U.S. Army or the U.S. Government. The U.S. Government is authorized to reproduce and distribute reprints for Government purposes notwithstanding any copyright notation herein.
Emails: 
\{\href{mailto:samuel.rey.escudero@urjc.es}{samuel.rey.escudero}, \href{mailto:andrei.buciulea@urjc.es}{andrei.buciulea}, \href{mailto:antonio.garcia.marques@urjc.es}{antonio.garcia.marques}\}@urjc.es, \{\href{mailto:nav@rice.edu}{nav},  \href{mailto:segarra@rice.edu}{segarra}\}@rice.edu.
}%
}
\address{$^*$ Dept. of Signal Theory and Communications, King Juan Carlos University, Madrid, Spain \\
$^\dagger$ Dept. of Electrical and Computer Engineering, Rice University, Houston, USA}
\maketitle

\newcommand{\mad}[1]{{\color[RGB]{218,112,214}[\textbf{Mad:} #1]}}
\newcommand{\sam}[1]{{\color{blue}[\textbf{Sam:} #1]}}
\newcommand{\ab}[1]{{\color[RGB]{204,163,0}[\textbf{AB:} #1]}}

\begin{abstract}
We consider \textit{fair network topology inference} from nodal observations.
Real-world networks often exhibit biased connections based on sensitive nodal attributes.
Hence, different subpopulations of nodes may not share or receive information equitably.
We thus propose an optimization-based approach to accurately infer networks while discouraging biased edges.
To this end, we present bias metrics that measure topological demographic parity to be applied as convex penalties, suitable for most optimization-based graph learning methods.
Moreover, we encourage equitable treatment for any number of subpopulations of differing sizes.
We validate our method on synthetic and real-world simulations using networks with both biased and unbiased connections.
\end{abstract}



\vspace{-.15cm}

\section{Introduction}
\label{s:intro}

\vspace{-.15cm}

Many types of data benefit from or require relational information for practical tasks.
For example, brain networks greatly improve neuronal analysis by statistically connecting behavior between neurons~\cite{chang2022lowrank}.
For social and communication networks and many other systems, understanding relationships between entities is the end rather than the
means~\cite{stoica2018algorithmic}.
Systems modeled by graphs permit evaluation based not only on characteristics of data but also their complex interactions~\cite{kolaczyk2009book}.

The increasing application of graph-based tools warrants care for their broader effects.
This includes fairness, which aims to prevent methods that discriminate based on sensitive attributes such as gender, race, or age~\cite{stoica2018algorithmic}.
An important yet nascent branch of fairness concerns graphs modeling biased relationships, that is, preferential attachments between certain subpopulations~\cite{dai2021say,li2021dyadic}.
Indeed, discriminatory connections are known to exacerbate bias when information propagates over a network~\cite{dai2021say}.
Recent works attempt to ameliorate the effect of biased connections, primarily for downstream tasks~\cite{bose2019compositional,rahman2019fairwalk,agarwal2021towards,kleindessner2019guarantees}, but directly addressing imbalanced connectivity is limited~\cite{li2021dyadic}.
Moreover, it is typically assumed that the underlying network is known, a frequently unrealistic scenario~\cite{tarzanagh2023fair,zhang2023unified}.

We propose \textit{fair network topology inference} from observations on nodes belonging to different groups.
Learning graphs from data is a prevalent task, boasting several methods based on different relationships between observations and network topology~\cite{mateos2019connecting}.
Examples include correlation networks, Gaussian Markov random fields (GMRFs)~\cite{friedman2008sparse,meinshausen06}, graph signal processing (GSP) methods~\cite{segarra2017network,kalofolias2016learn,buciulea2022learning}, and those with more complex priors~\cite{rey2022enhanced,roddenberry2021network,navarro2022jointb}.
However, when the underlying graph discriminates by nodal groups, these methods may return networks with biased connections.
Thus, our goal is to infer graphs that accurately explain nodal observations while reducing edge favoritism toward certain pairs of groups. 

We present a flexible optimization-based method that recovers a network from nodal observations while encouraging balanced edges across groups.
To the best of our knowledge, ours is the first method to \textit{consider fairness explicitly for estimating graphs from data.}
To this end, we offer two topological bias metrics for use as penalties to mitigate unfair connections.
We exemplify them assuming that nodal observations are \textit{stationary}~\cite{segarra2017network}, a model that subsumes correlation networks, GMRFs, and others.
Our contributions are as follows.

\vspace{-.15cm}

\begin{itemize}[labelwidth=1em,leftmargin =\dimexpr\labelwidth+\labelsep\relax]
    \setlength\itemsep{-.05cm}
    \item[1)] We formalize group fairness for network topology inference by introducing two topological bias metrics.
    \item[2)] We present a flexible optimization-based approach to recover networks while discouraging biased connections.
    \item[3)] We provide theoretical results for the uniqueness of the estimated graph based on previous work in~\cite{segarra2017network}.
    \item[4)] We validate our approach empirically using both synthetic and real-world data.
\end{itemize}

\section{Preliminaries}
\label{s:bg}

In this section, we provide necessary definitions and background for our problem of interest.
We consider undirected graphs of the form $\ccalG=(\ccalV,\ccalE)$ consisting of a set of $N$ nodes $\ccalV$ and edges $\ccalE\subseteq\ccalV\times\ccalV$.
We let $\bbA\in\reals^{N\times N}$ denote the adjacency matrix of $\ccalG$, where $A_{ij}=A_{ji}\neq 0$ if and only if $(i,j)\in\ccalE$.
The adjacency matrix $\bbA$ is an example of a graph shift operator (GSO) $\bbS$, which encodes the support of the graph~\cite{sandryhaila2013discrete}, with other choices including the graph Laplacian.
We represent nodal observations as \textit{graph signals} $\bbx\in\reals^N$, where entry $x_i$ corresponds to a signal observed on the $i$-th node.
Each node belongs to one of $G$ groups, represented by the indicator matrix $\bbZ\in\{0,1\}^{N\times G}$, where $Z_{ig}=1$ if node $i$ belongs to group $g$, otherwise $Z_{ig}=0$.
We let $\bbz_g$ denote the $g$-th column of $\bbZ$ and $N_g$ the number of nodes in group $g$.

\vspace{.1cm}

\noindent\textbf{Graph stationarity}.
Network topology inference requires a formal relationship between graph signals and their underlying network~\cite{mateos2019connecting}.
To exemplify our approach, in this paper we consider that the graph signals are \textit{stationary} on $\ccalG$~\cite{segarra2017network,djuric2018cooperative}.
Graph stationarity, which has been used previously in the context of graph learning \cite{segarra2017network}, implies that the covariance of the graph signals $\bbC = \mbE[\bbx\bbx^\top]$ 
can be written as a polynomial of the GSO $\bbS$. 
As such, graph stationarity is a fairly general model encompassing correlation networks and GMRFs as particular cases. 
An implication of stationarity is that the graph signal covariance matrix commutes with the GSO, that is, $\bbC\bbS = \bbS\bbC$, a key fact for our graph learning algorithm.

\vspace{.1cm}

\noindent\textbf{Fairness on graphs}.
Notions of group fairness such as demographic parity (DP) require equitable treatment regardless of group membership~\cite{hardt2016equality}.
For example, recommendation systems are deemed fair if sensitive demographic information is not considered~\cite{bose2019compositional,rahman2019fairwalk}.
Previous works for fairness on graphs include mitigating the effect of biased connections on node embeddings~\cite{bose2019compositional,rahman2019fairwalk,agarwal2021towards}, link prediction~\cite{bose2019compositional,rahman2019fairwalk}, and graph clustering~\cite{kleindessner2019guarantees,tarzanagh2023fair,zhang2023unified}.
While these works consider the group fairness of edges, they primarily aim to reduce biases in downstream tasks~\cite{li2021dyadic}.
As an example, the bias metric in~\cite{kose2023fairnessaware} measures the correlation between nodal groups and network topology, but their goal is to design fair graph filters to remove biases from data.
Most assume that the graph is completely known, excepting a few fair graph \textit{clustering} methods~\cite{tarzanagh2023fair,zhang2023unified}.
Both jointly infer the graph and cluster nodes to yield unbiased clusters, but the learned graph is not explicitly encouraged to have fair connections. 
The most relevant work to our own is~\cite{kose2024fairwire}, which proposes a fair generative model for \textit{random synthetic} graphs.
While they also promote balanced edges in the resultant graphs, our task is fundamentally different as we estimate networks to explain observed data on nodes.


\section{Topological demographic parity}
\label{s:dp}


Group fairness, which includes DP and equality of odds~\cite{hardt2016equality}, considers an algorithm fair if its decisions are group-agnostic.
For \textit{dyadic} group fairness, topological DP requires that edges exhibit no preference for connecting the same group or different groups.
In our case, we define DP as
\begin{equation*}
    \mbP[ (i,j)\in\ccalE | Z_{ig}=Z_{jg}=1 ] = 
    \mbP[ (i,j)\in\ccalE | Z_{ig}=Z_{jh}=1 ]
\end{equation*}
for every $g,h\in\{1,2,\dots,G\}$.
When DP is not achieved, we compute the empirical \textit{gap in DP} as
\vspace{-0.2cm}
\begin{equation}\label{eq:dp_gap}
    \vspace{-0.2cm}
    \dpgap(\bbA,\bbZ) := 
    \sum_{g=1}^G \sum_{h\neq g}
    \left| \frac{ \bbz_g^\top \bbA \bbz_g }{N_g^2-N_g}
    - \frac{ \bbz_g^\top \bbA \bbz_{h} }{N_g N_{h}} \right|.
\end{equation}
Minimizing $\dpgap$ encourages balanced numbers of edges connecting nodes in the same group or different groups.
The metric in~\cite{kose2024fairwire} is similar to 
$\dpgap$. However, they balance within- and across-group edges \textit{averaged over other groups}, while we require balance between every pair of groups.

Despite its originally different purpose, we adapt the bias metric in~\cite{kose2023fairnessaware} for an alternate measure of DP gap and extend it beyond $G=2$ groups.
Recall that~\cite{kose2023fairnessaware} proposes a correlation metric between nodal groups and network topology.
Consider the matrix $\bbB\in\reals^{G\times N}$ such that $B_{gi} = (G-1)/N_g$ if node $i$ belongs to group $g$, and $B_{gi} = -1/N_{h}$ if node $i$ belongs to a different group, $h\neq g$.
We present our \textit{nodewise} gap in DP,
\begin{equation}\label{eq:dp_node}
    \dpngap(\bbA,\bbZ) := 
    \|\bbB\bbA\|_1,
\end{equation}
where a high value of $\dpngap$ implies a high correlation between the connectivity of the $i$-th node and a particular group.
Thus, node preferences for certain groups are reflected in high values of $\dpngap$.
Note that our metric $\dpngap$ differs from that of~\cite{kose2023fairnessaware} as theirs measures biased aggregations of node features instead of imbalanced edge connectivity across group pairs.
For further intuition, observe that
\vspace{-0.1cm}
\begin{alignat}{3}&
     \|\bbB\bbA\|_1 = 
    \sum_{i=1}^N \sum_{g=1}^G
    \left| 
    \sum_{h\neq g}
    \frac{[\bbA\bbz_g]_i}{N_g} -
    \frac{[\bbA\bbz_{h}]_i}{N_{h}}
    \right|,
    \nonumber
&\end{alignat}
which increases if any node is more likely to connect to certain groups.
Thus, measuring the correlation between groups and topology via $\dpngap$ is analogous to measuring fair connections \textit{per node}.
This nodewise metric is more restrictive than the groupwise $\dpgap$, which permits more flexible connections as long as they are balanced across group pairs.
The next section incorporates $\dpgap$ and $\dpngap$ as penalties in optimization-based graph learning problems to mitigate biased connections for network topology inference.

\begin{figure*}[!t]
    \centering
    \begin{minipage}[b]{.62\textwidth}
        \centering
        \begin{subfigure}{.48\textwidth}
            \centering
            \scalebox{.88}{\begin{tikzpicture}[baseline,scale=1.02,trim axis left, trim axis right]

\pgfplotstableread{data/edgeratio.csv}\errtable

\definecolor{btr}{RGB}{255,255,255}
\definecolor{bnp}{RGB}{33, 102, 172}
\definecolor{bdp}{RGB}{67, 147, 195}
\definecolor{bnw}{RGB}{146, 197, 222}
\definecolor{fnp}{RGB}{178, 24, 43}
\definecolor{fdp}{RGB}{214, 96, 77}
\definecolor{fnw}{RGB}{244, 165, 130}

\definecolor{fax}{RGB}{200,0,3}
\definecolor{bax}{RGB}{4,50,255}

\begin{axis}[
    xlabel={(a) Across-group edge ratio},
    xmin=0.125,
    xmax=0.875,
    ylabel={Bias},
    ymin=0.05,
    ymax=1.7,
    grid style=densely dashed,
    grid=both,
    legend style={
        at={(1,1)},
        anchor=north east,
        font=\footnotesize},
    legend columns=2,
    width=185,
    height=160,
    xtick pos=left,
    ytick pos=left,
    label style={font=\small},
    tick label style={font=\small}
    ]
    
    \addplot[black, mark=x, densely dotted] table [x=Edge_Ratio, y=Bias_True] {\errtable};
    \addplot[blue!80!black, mark=o, solid] table [x=Edge_Ratio, y=Bias_NTI] {\errtable};
    \addplot[blue!65!white, mark=o, solid] table [x=Edge_Ratio, y=Bias_DP] {\errtable};
    \addplot[blue!25!white, mark=o, solid] table [x=Edge_Ratio, y=Bias_NW] {\errtable};

    \addlegendentry{True}
    \addlegendentry{None}
    \addlegendentry{$\Delta\mathrm{DP}$}
    \addlegendentry{$\Delta\mathrm{DP}_\mathrm{node}$}
\end{axis}

\end{tikzpicture}}
        \end{subfigure}
        \hfill
        \begin{subfigure}{.48\textwidth}
            \centering
            \scalebox{.88}{\begin{tikzpicture}[baseline,scale=1.02,trim axis left, trim axis right]

\pgfplotstableread{data/edgeratio.csv}\errtable

\definecolor{btr}{RGB}{255,255,255}
\definecolor{bnp}{RGB}{33, 102, 172}
\definecolor{bdp}{RGB}{67, 147, 195}
\definecolor{bnw}{RGB}{146, 197, 222}
\definecolor{fnp}{RGB}{178, 24, 43}
\definecolor{fdp}{RGB}{214, 96, 77}
\definecolor{fnw}{RGB}{244, 165, 130}

\definecolor{fax}{RGB}{200,0,3}
\definecolor{bax}{RGB}{4,50,255}

\begin{axis}[
    xlabel={(b) Across-group edge ratio},
    xmin=0.125,
    xmax=0.875,
    ylabel={Error},
    ymin=0.05,
    ymax=0.35,
    grid style=densely dashed,
    grid=both,
    legend style={
        at={(0,1)},
        anchor=north west,
        font=\footnotesize},
    legend columns=1,
    width=185,
    height=160,
    xtick pos=left,
    ytick pos=left,
    label style={font=\small},
    tick label style={font=\small},
    mark options={rotate=45, scale=1.2}
    ]
    
    \addplot[red!40!black, mark=square, solid] table [x=Edge_Ratio, y=Frob_NTI] {\errtable};
    \addplot[red!90!black, mark=square, solid] table [x=Edge_Ratio, y=Frob_DP] {\errtable};
    \addplot[red!50!white, mark=square, solid] table [x=Edge_Ratio, y=Frob_NW] {\errtable};

    \addlegendentry{None}
    \addlegendentry{$\Delta\mathrm{DP}$}
    \addlegendentry{$\Delta\mathrm{DP}_\mathrm{node}$}
\end{axis}

\end{tikzpicture}}
        \end{subfigure}
        \vspace{-.1cm}
        \caption{\small{
        (a) Bias of true and estimated networks, measured as $\dpgap$ divided by edge density.
        (b) Error of estimates, measured as $\| (\bbA/\|\bbA\|_F) - (\hbA/\|\hbA\|_F) \|_F^2/2$.
        }
    \label{fig:synth_exps}}
    \end{minipage}
    \hfill
    \begin{minipage}[b]{.34\textwidth}
        \footnotesize
        \centering
        \begin{tabular}{m{2.6em} m{9.6em} m{1.8em} m{2.0em}}
        \hline
            &
            \multicolumn{1}{c}{\bf Method} & 
            \multicolumn{1}{c}{\bf Bias} &
            \multicolumn{1}{c}{\bf Error} \\
        \hline
            \multirowcell{6}{Unfair \\ graph} & True & 1.214 & \textemdash \\
            & None ($\beta=0$) & 0.962 & 0.040 \\
            \hhline{~---}
            & $\dpgap$ ($\beta=100$) & 0.874 & 0.055 \\
            & $\dpngap$ ($\beta=100$) & 0.886 & 0.053 \\
            \hhline{~---}
            & $\dpgap$ ($\beta=1000$) & 0.705 & 0.091 \\
            & $\dpngap$ ($\beta=1000$) & 0.725 & 0.108 \\
            \hline
            \multirowcell{6}{Fair \\ graph} & True & 0.216 & \textemdash \\
            & None ($\beta=0$) & 0.197 & 0.040 \\
            \hhline{~---}
            & $\dpgap$ ($\beta=100$) & 0.192 & 0.042 \\
            & $\dpngap$ ($\beta=100$) & 0.194 & 0.050 \\
            \hhline{~---}
            & $\dpgap$ ($\beta=1000$) & 0.164 & 0.063 \\
            & $\dpngap$ ($\beta=1000$) & 0.177 & 0.129 \\
        \end{tabular}
        \vspace{-.1cm}
        \captionof{table}{\small{
        Bias and error of estimated networks for fair versus unfair group labels.
        }\label{tab:sbm}}
    \end{minipage}
    \vspace{-.4cm}
\end{figure*}

\section{Fair network topology inference}
\label{s:method}

We infer \textit{fair} network topology from \textit{stationary} graph signals as described in Section~\ref{s:bg}~\cite{segarra2017network}.
Given a set of $M$ observations collected in the matrix $\bbX\in\reals^{N\times M}$, we estimate the sample covariance matrix $\hbC= \frac{1}{M}\bbX\bbX^\top$ to obtain an adjacency matrix $\hbA$ that approximately commutes with $\hbC$ via the following nonconvex optimization problem
\vspace{-0.1cm}
\begin{alignat}{3}&
    \hbA = \argmin_{\bbA} \|\bbA\|_0 + \beta \Delta\mathrm{DP}(\bbA,\bbZ) 
    &\nonumber\\&
    \st ~~ \|\bbA\hbC - \hbC\bbA\|_F \leq \epsilon, ~~\bbA\in\ccalA,
    \label{eq:nti_stat}
&\end{alignat}
where $\beta>0$ determines the tradeoff between sparsity and DP.
The first term in the objective promotes sparsity, a common desirable structural characteristic for parsimonious estimates and reduced downstream computation.
Importantly,~\eqref{eq:nti_stat} serves as a highly flexible framework.
First, the second term in~\eqref{eq:nti_stat} can be any fairness metric, including $\dpgap$ or $\dpngap$.
Second, the chosen penalty can suit any optimization-based approach such as graphical lasso~\cite{friedman2008sparse}.
We emphasize stationary graph signals in this manuscript; hence, the first constraint recovers a graph on which observations $\bbX$ are approximately stationary, controlled by $\epsilon\geq 0$.
However, the journal version of this paper will extend our approach to other common graph signal models, including GMRFs.
The set $\ccalA$ denotes the set of valid, nontrivial adjacency matrices.
While we estimate the adjacency matrix $\bbA$, our approach can easily accommodate other GSOs that can be specified by $\ccalA$.

As the $\ell_0$ norm is nonconvex, a common approach is to replace it with a convex relaxation, 
\vspace{-.1cm}
\begin{alignat}{3}&
    \tbA = \argmin_{\bbA} \|\bbA\|_1 + \beta \Delta\mathrm{DP}(\bbA,\bbZ) 
    &\nonumber\\&
    \st ~~ \|\bbA\hbC - \hbC\bbA\|_F \leq \epsilon, ~~\bbA\in\ccalA,
    \label{eq:nti_stat_cvx}
&\end{alignat}
where we merely replace the nonconvex $\ell_0$ norm with the $\ell_1$ norm, whose convexity yields efficient solutions and theoretical guarantees.
Not only do we enjoy the benefits of a convex objective, but we can also provide conditions that guarantee obtention of the sparsest solution, that is $\tbA = \hbA$.

The rest of this section is devoted to providing theoretical guarantees for \eqref{eq:nti_stat_cvx}. 
We restrict our result to the case where $\epsilon = 0$, but, building on the approach in~\cite{navarro2023joint}, we can extend the result to $\epsilon>0$.
We next show sufficient conditions for $\hbA$ to be equivalent to $\tbA$.
First, we require additional definitions for the statement of our result.
We let $\hba = \mathrm{vec}(\hbA)$ and $\tba = \mathrm{vec}(\tbA)$ respectively denote the concatenation of columns of $\hbA$ and $\tbA$.
Then, there exist matrices $\bbPsi$ and $\bbPhi$ and a vector $\bbb$~\cite{segarra2017network,navarro2022joint,navarro2023joint} such that~\eqref{eq:nti_stat_cvx} can be rewritten as
\begin{alignat}{3}&
    \tba = \argmin_{\bba} \|\bbPsi\bba\|_1 ~~\st~~ \bbPhi\bba = \bbb.
    \label{eq:nti_stat_vec}
&\end{alignat}
Let $\ccalI = \mathrm{supp}(\hba)$ and $\ccalJ = \mathrm{supp}(\bbPsi\hba)$, where $\mathrm{supp}(\bbx)$ denotes the support of the vector $\bbx$.
For a matrix $\bbX$, let $\bbX_{\ccalI,\cdot}$ and $\bbX_{\cdot,\ccalI}$ be the submatrices of $\bbX$ consisting of rows and columns selected by $\ccalI$, respectively.
Moreover, let $\ccalJ^c$ be the complement of the set $\ccalJ$.
Our theoretical result is as follows.

\vspace{-.1cm}

\begin{theorem}\label{thm:cvx}
    Assume that~\eqref{eq:nti_stat} and~\eqref{eq:nti_stat_cvx} are feasible. Then, $\hbA = \tbA$ if the following two conditions are satisfied:
    \vspace{-.1cm}
    \begin{itemize}[labelwidth=1em,leftmargin =\dimexpr\labelwidth+\labelsep\relax]
        \item[1)] $\bbPhi_{\cdot,\ccalI}$ is full column rank, and
        \vspace{-.2cm}
        \item[2)] There exists a constant $\psi>0$ such that
        \vspace{-.1cm}
        \be
            \| \bbPsi_{\ccalJ^c,\cdot}(\psi^{-2} \bbPhi^\top\bbPhi + \bbPsi_{\ccalJ^c,\cdot}^\top \bbPsi_{\ccalJ^c,\cdot} )^{-1} \bbPsi_{\ccalJ,\cdot} \|_{\infty} < 1.
        \ee
    \end{itemize}
\end{theorem}

\vspace{-.1cm}

\begin{myproof}[of Theorem~\ref{thm:cvx}]
    Despite the novel penalty, the convex objective in~\eqref{eq:nti_stat_cvx} can be vectorized similarly to~\cite{navarro2022joint}. 
    Hence, we impose conditions analogous to those from~\cite[Theorem 1]{navarro2022joint}, albeit with different topological implications.
    The result then follows from~\cite[Theorem 1]{navarro2022joint}.
\end{myproof}

The conditions of Theorem~\ref{thm:cvx} guarantee that $\hbA$ is the unique solution to~\eqref{eq:nti_stat_cvx}.
More specifically, condition \textit{1)} ensures that the solution to~\eqref{eq:nti_stat_cvx} is unique, while condition \textit{2)} provides a dual certificate verifying that $\hbA$ is indeed a feasible solution to~\eqref{eq:nti_stat_cvx}.
Thus, under the conditions of Theorem~\ref{thm:cvx}, we are guaranteed to obtain the desired sparse solution $\hbA$ with the computationally friendly convex relaxation~\eqref{eq:nti_stat_cvx}.

\begin{figure*}[!t]
    \centering
    \begin{minipage}[!m]{.26\textwidth}
        \centering
        \includegraphics[width=\textwidth]{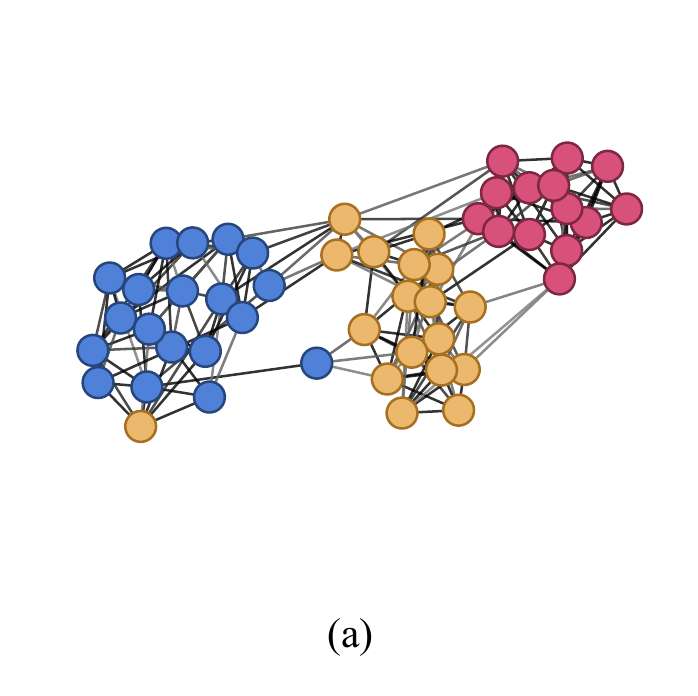}
    \end{minipage}
    \begin{minipage}[!m]{.26\textwidth}
        \centering
        \includegraphics[width=\textwidth]{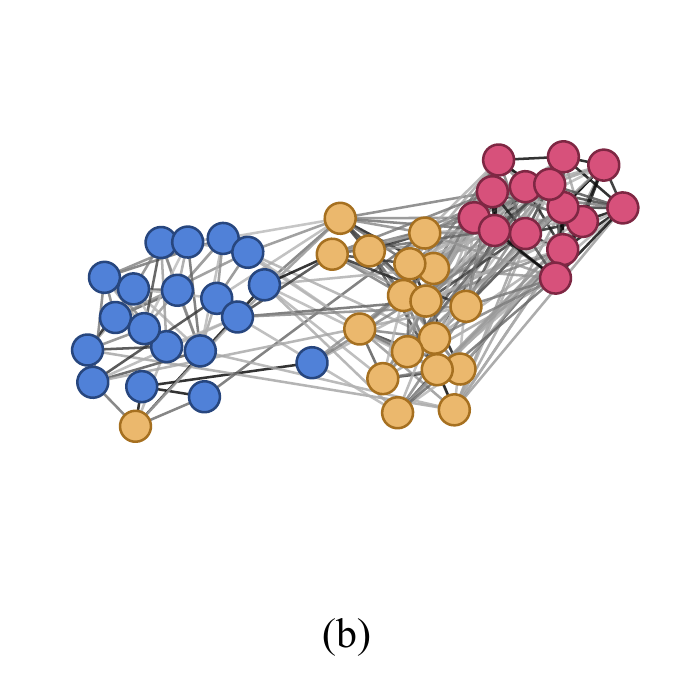}
    \end{minipage}
    \hspace{.75cm}
    \begin{minipage}[!b]{.27\textwidth}
        \centering
        \scalebox{.9}{\begin{tikzpicture}[baseline,scale=1.02,trim axis left, trim axis right]

\pgfplotstableread{data/senate113_bias.csv}\biastable
\pgfplotstableread{data/senate113_comm.csv}\commtable


\begin{semilogxaxis}[
    xlabel={(c) Fair penalty weight $\beta$},
    xmin=1e2,
    xmax=1e5,
    axis y line*=left,
    ylabel={Bias},
    ymin=-0.02,
    ymax=0.22,
    ytick={0, 0.05, 0.1, 0.15, 0.2},
    yticklabel style={
        /pgf/number format/fixed,
        /pgf/number format/precision=3
    },
    grid style=densely dashed,
    grid=both,
    legend style={
        at={(0,1)},
        anchor=north west,
        font=\footnotesize},
    legend columns=1,
    width=185,
    height=160,
    xtick pos=left,
    ytick pos=left,
    label style={font=\small},
    tick label style={font=\small}
    ]

    \addplot[blue!80!black, mark=o, solid] table [x=beta, y=nti] {\biastable};
    \addplot[blue!65!white, mark=o, solid] table [x=beta, y=fnti] {\biastable};
    \addplot[blue!25!white, mark=o, solid] table [x=beta, y=nfnti] {\biastable};

\end{semilogxaxis}

\begin{loglogaxis}[
    xlabel={},
    xtick={},
    xmin=1e2,
    xmax=1e5,
    ylabel={Error},
    ytick={1e-6, 1e-5, 1e-4, 1e-3, 1e-2},
    ylabel shift=-10pt,
    ymin=3.98107171e-7,
    ymax=0.02511886431,
    axis y line*=right,
    legend style={
        at={(1,1)},
        anchor=north east,
        font=\footnotesize},
    legend columns=1,
    width=185,
    height=160,
    label style={font=\small},
    tick label style={font=\small}, 
    mark options={rotate=45, scale=1.2}
    ]


    \addplot[red!40!black, mark=square, solid] table [x=beta, y=nti] {\commtable};
    \addplot[red!90!black, mark=square, solid] table [x=beta, y=fnti] {\commtable};
    \addplot[red!50!white, mark=square, solid] table [x=beta, y=nfnti] {\commtable};

\end{loglogaxis}

\end{tikzpicture}}
    \end{minipage}
    \vspace{-.2cm}
    \caption{\small{
    (a) Estimated senate network \textit{without bias mitigation} for 113th congress with edges colored by weight.
    (b) Estimated senate network with $\dpgap$ for 113th congress with edges colored by weight.
    (c) Bias and commutativity error of estimated senate networks as tuning parameter $\beta$ increases. 
    Error is computed as $\|\hbC\hbA - \hbA\hbC\|_F / \|\hbC\|_F$ for an estimated network $\hbA$.
    Labels from the legend in Fig.~\ref{fig:synth_exps} apply.
    }}
    \label{fig:real_exps}
    \vspace{-.4cm}
\end{figure*}

\section{Numerical evaluation}
\label{s:results}
Due to space limitations, we present here some of the experiments and refer interested readers to the GitHub repository (\url{https://github.com/mn51/fair_nti}) for full details on the simulation setup and additional experiments.

\vspace{.1cm}

\noindent\textbf{Varying network topology.}
We first assess~\eqref{eq:nti_stat} as topology varies in Figs.~\ref{fig:synth_exps}a and \ref{fig:synth_exps}b.
We consider $N=30$ nodes and two equal-sized groups $G=2$, where group assignments and the number of edges remain fixed.
We vary the ratio of edges between groups, where the underlying graph ranges from a high within-group edge preference to a high across-group edge preference.
The true graph is generated by first creating a connected component per group and then rewiring an increasing subset of within-group edges to connect nodes across groups.
Figs.~\ref{fig:synth_exps}a and \ref{fig:synth_exps}b respectively show the bias and error as we increase the across-group edge ratio for three approaches: (i) ``True'', for the bias of the true graph, (ii) ``None'',~\eqref{eq:nti_stat} for $\beta=0$, (iii) ``$\dpgap$'',~\eqref{eq:nti_stat} as stated, and (iv)~``$\dpngap$'', where $\dpngap$ replaces $\dpgap$ in~\eqref{eq:nti_stat}.

For the bias of method ``True'' in Fig.~\ref{fig:synth_exps}a, the fairest setting occurs when the across-group ratio is 0.5, as expected for an equal number of within- and across-group edges.
Notably, estimations from ``$\dpgap$'' and ``$\dpngap$'' show similar levels of bias, superior to ``None''.
However, in~Fig.~\ref{fig:synth_exps}b ``$\dpgap$'' achieves a lower error than ``$\dpngap$'', which is a stronger fairness metric (see Section~\ref{s:dp}).
Indeed, when the underlying graph is fair at an across-group edge ratio of 0.5, errors are closely aligned for ``$\dpgap$'' and ``None'' but not ``$\dpngap$'', which aims to ensure that every node balances connections to all groups.
Moreover, note that a larger gap in bias between true and estimated graphs is generally proportional to error as the true graph becomes more unfair.
We thus demonstrate that both penalties $\dpgap$ and $\dpngap$ effectively reduce bias as the topology varies in fairness, while $\dpgap$ maintains lower error than $\dpngap$ particularly when the true graph is fair.

\vspace{.1cm}

\noindent\textbf{Varying group assignments.}
In this experiment, we test the effect of $\dpgap$ and $\dpngap$ when nodes are assigned to $G=2$ groups either fairly or unfairly.
We generate a fixed 2-community graph of $N=30$ nodes and consider two settings: In the fair case, we assign nodes to groups uniformly at random, while in the unfair case, group assignments coincide with the communities.
Table~\ref{tab:sbm} presents the bias and error for the fair and unfair settings averaged over 100 graph realizations.
We evaluate the same three approaches as the previous experiment for fairness promoting weights $\beta\in\{100,1000\}$.

For the unfair graph, ``$\dpgap$'' and ``$\dpngap$'' experience a noticeable increase in error, particularly for the larger $\beta=1000$.
This is expected as~\eqref{eq:nti_stat} attempts to obtain a fair estimate even though the true graph has imbalanced connections.
Conversely, when the true graph is fair, ``$\dpgap$'' and ``$\dpngap$'' generally maintain a lower error, as encouraging fairness aligns with the setting of the true graph.
Moreover, ``$\dpgap$'' is robust in the fair case as the error for $\beta=1000$ does not increase greatly even with a larger decrease in bias.
However, the stronger penalty $\dpngap$ suffers greater error for larger $\beta$, even when the graph is fair.
Thus, our fairness metrics are not only well suited to estimating networks with unbiased connections, but we can also maintain low estimation error while improving bias.

\vspace{.1cm}

\noindent\textbf{Real-world senate data.}
Finally, we evaluate each method for a real-world dataset of U.S. Congress roll-calls~\cite{lewis2020votes}.
We consider the $M=657$ votes of the 113th Congress (2013 to 2015) across $N=51$ nodes, which represent the opinion of the President and those of the 50 states.
As every state houses two senators, nodal observations for each state are the sum of the votes for both senators, where yea, nay, and other choices such as abstention are respectively represented by 1, -1, and 0.
Group labels correspond to political party, where each node can be Democratic or Republican if both senators belong to the respective party or Mixed if their parties differ.
We estimate pairwise state connections while varying the parameter $\beta$ to observe the effect of bias mitigation on the resultant senate networks.
In Fig.~\ref{fig:real_exps}c, we present the estimated bias and error as $\beta$ grows.
As we do not have a true graph, we instead evaluate how well our estimates satisfy the assumed graph signal model.
Thus, our error in Fig~\ref{fig:real_exps}c measures the violation in commutativity between $\hbC$ and the estimated adjacency matrices.
A low error implies that we can obtain networks on which the observed votes are approximately stationary.

Observe that as politicians from the same party tend to exhibit similar voting behavior, ``None'' yields biased networks.
Both ``$\dpgap$'' and ``$\dpngap$'' are able to reduce connection biases while maintaining a relatively low commutativity error.
Moreover, while bias of the stricter ``$\dpngap$'' decreases faster with respect to $\beta$, ``$\dpgap$'' achieves a consistently lower error than ``$\dpngap$'' for the same bias level.
Figs.~\ref{fig:real_exps}a and~\ref{fig:real_exps}b respectively visualize estimates from ``None'' and ``$\dpgap$''.
The fairer graph, depicted in Fig.~\ref{fig:real_exps}b, shows an increase in connections across parties while still adhering to the commutativity constraint.
The added across-group edges identify which senators from different parties demonstrate the most similar voting behavior, suggesting potential collaborations.
Hence, our modified network inference method obtains realistic yet unbiased graphs in real-world settings.

\section{Conclusion}
\label{s:conclusion}

In this work, we presented inference of unbiased networks from stationary graph signals, where nodes are partitioned into groups according to a discrete sensitive attribute.
We introduced two bias metrics as optimization penalties that measure the gap in DP for edges connecting different pairs of groups.
We then proposed an optimization problem for recovering a network that approximately satisfies both stationarity and a desired balance of edges between groups.
While we emphasized DP and stationarity, the proposed optimization problem functions as a framework for any graph signal model and any topological bias metric.
In the extension of this work, we expand on both the theoretical tradeoff between accuracy and fairness and also comparisons of other popular graph models and notions of fair connectivity.

\bibliographystyle{ieeetr}
\bibliography{citations}

\end{document}